\documentclass[%
 reprint,
 amsmath,amssymb,
 aps,
 pra,
]{revtex4-2}

\usepackage{graphicx}
\usepackage{dcolumn}
\usepackage{bm}

\usepackage{siunitx}
\usepackage{float}
\usepackage[caption = false]{subfig}

\begin{document}

\preprint{APS/123-QED}

\title{Theoretical Study of the Enhancement of Light Saturation Phenomena of Krypton at Critical Ionization Photon Energies}
\thanks{manuscript draft version}%

\author{Jiaxin Ye}%
\author{Chen Yang}%
 \email{yangchen@scu.edu.cn}
\author{Gang Jiang}%
 \email{gjiang@scu.edu.cn}
\affiliation{%
 Institute of Atomic and Molecular Physics, Sichuan University, Chengdu 610065, China\\
 Key Laboratory of High Energy Density Physics and Technology, Ministry of Education, Chengdu 610065, China
}%


\author{Yixuan Yang}
\affiliation{
College of Electrical Engineering, Sichuan University, Chengdu 610065, China
}%


\date{\today}

\begin{abstract}
By calculating the correlation between the total photoionization cross-section of the ground state of the Kr atom and photon energy, three particular photon energies close to the near inner orbital energy of 1.75 keV, 1.90 keV, and 14.30 keV are determined in this work. The dynamical simulation under 17.50 keV photon energy in the experimental conditions is achieved by implementing the Monte Carlo method and optimizing the photon flux modeling parameters. As a result, our calculated data are more consistent with the experimental phenomena. The light saturation phenomenon of Kr at 1.75 keV, 1.90 keV, 14.30 keV, and 17.50 keV energies is further calculated and researched using the optimized photon flux model theory. We statistically compare the main ionization paths under those four specific photon energies and calculate the population changes of various hollow atoms. The results demonstrate that the ratio of hollow atoms produced at the critical ionization photon energy is high. Furthermore, the change of position is smooth, showing the significant difference between the generation mode of ions with low photon energy and those with high photon energy, which has important reference significance for studying hollow atoms with medium and high charge states.
\end{abstract}

\maketitle


\section{Introduction}

X-ray free-electron laser (XFEL) is known as the ``fourth-generation light source.''\cite{Pellegrini2016} XFEL opens up a novel window in exploring substances at the atomic and molecular scales of space and time, photographing chemical processes within a few milliseconds or less, and revealing the structure and dynamics of complex molecular systems.\cite{sobolev_megahertz_2020,lehmkuhler_emergence_2020} In recent years, free-electron laser measurements of protein structures have become an increasingly valuable tool in biology,\cite{pellegrini_history_2012,caleman_ultrafast_2015}. At the same time, coherent X-ray diffraction imaging provides images of small structures, which is a major focus of crystallography and protein research.\cite{kobayashi_dark-field_2014} It is of great significance to investigate the dynamical evolution of atoms under a laser in order to assess the radiation damage phenomenon in sample structure detection using free-electron lasers.\cite{ho_resonance-mediated_2015,howells_assessment_2009,doumy_nonlinear_2011} 

In experiments, X-ray photoionization can produce light transparency in aluminum.\cite{bob_nagler_etal_turning_2009} When an atom interacts with the XFEL, the inner electrons are ionized, resulting in light saturation, reducing the probability of light-induced damage, and the diffraction image obtained achieves the best quality.\cite{deng_charge-state_2019,feng_lei_2000_2017,yoneda_saturable_2014,son_impact_2011} Due to the limitation of photon energy generated by a laser and the phenomenon of hollow atoms in medium- and high-$Z$ element atoms have the high decay probability, the light saturation phenomenon is not apparent, the research on medium- and high-$Z$ element hollow atoms is relatively scarce. Nevertheless, if the photon energy is close to the ionization orbital energy, when the orbital ionization energy increases and exceeds the photon energy after the inner electron is ionized, it is possible to make the phenomenon of light saturation evident and durable, which makes the practical application of hollow atoms in the characterization of material structure possible. 

The photon energy close to the orbital energy is called the critical ionization photon energy (CIPE). Consequently, we extend the study of hollow atoms to Kr and examine the phenomenon of hollow atoms at photon energies close to their orbital energy. In this work, we performed a genetic algorithm (GA) to determine laser parameters in the model and simulate the dynamical evolution of the Kr atom under three CIPEs by theoretical calculation.

\section{Theoretical method}
\subsection{Theoretical model}
Our calculation of the photoionization cross-section is based on the \emph{ab initio} calculation framework of non-relativistic quantum electrodynamics and perturbation theory, using the Hartree-Fock-Slater model.\cite{santra_concepts_2009} The cross-section for photoionization of an electron in the $i$th occupied spin orbital follows as

\begin{eqnarray}
    \sigma_{i}=&&\frac{4 \pi^{2}}{\omega_{\mathrm{in}}} \alpha \sum_{f}\left|\left\langle\varphi_{f}\left|e^{\mathrm{i} \boldsymbol{k}_{\mathrm{in}} \cdot \boldsymbol{x}} \epsilon_{\boldsymbol{k}_{\mathrm{in}}, \lambda_{\mathrm{in}}} \cdot \frac{\nabla}{\mathrm{i}}\right| \varphi_{i}\right\rangle\right|^{2}\nonumber\\
    &&\times\delta\left(\varepsilon_{f}-\varepsilon_{i}-\omega_{\mathrm{in}}\right) ,
    \label{eq:1}
\end{eqnarray}
where $\omega_\text{in}$ is the incident photon energy, $\alpha$ is the fine structure constant, $\boldsymbol{x}$ is the position, $\epsilon_{\boldsymbol{k}_{\mathrm{in}}, \lambda_{\mathrm{in}}}$ is the polarization vector under the wave vector $\boldsymbol{k}$ and wavelength $\lambda$, $\varepsilon_f$ is the energy of unoccupied orbital, $\varepsilon_i$ is the energy of occupied orbital, $\varphi_f$ is the sum of unoccupied orbital wave functions, and $\varphi_i$ is the sum of occupied orbital wave functions.

\subsection{Photon flux modeling}

Since the calculation of photoionization probability depends on photon flux, the modeling of photon flux is crucial for us to fit the theoretical calculation results. In the actual experiment of free-electron laser, the distribution of light intensity meets not only the Gaussian distribution of time, but also the Gaussian distribution of position and photon frequency. We assume that the beam is a circular spot, and the general formula of light intensity satisfies the Gaussian distribution of space, time, and photon frequency,\cite{toyota_xcalib_2019,li_effects_2022,gao_single-_2016}

\begin{eqnarray}
    I_{\nu}(r, t)=&&I_{0} e^{-\ln 2 a\left(\frac{r_{1}}{\Delta}\right)^{2}} e^{-\ln 2\left(\frac{t}{T}\right)^{2}} \nonumber\\
    &&\times\sqrt{\frac{\ln 2}{\pi \Gamma^{2}}} e^{-\ln 2\left(\frac{h \nu-h \nu_{0}}{\Gamma}\right)^{2}}    .
\end{eqnarray}
For convenience, we define the relative radius position $r$ as

\begin{equation}
    r = \frac{r_1}{\Delta}  ,
\end{equation}
where $t$ is the time, $r_1$ is the relative center position, $\nu$ is the photon frequency, $\Delta$ is the radius, $\tau$ is the pulse half-width, $\Gamma$ is the half-width at the half-maximum (HWHM) of the photon frequency bandwidth, and $h\nu_0$ is the central photon energy. The exponential coefficient $a$ and the peak flux $I_0$ are the laser parameters to be calibrated. The photoionization rate can be obtained by integrating the photon frequency bandwidth,

\begin{equation}
    R_{i j}(r, t)=\int \frac{I_{\nu}(r, t)}{h \nu} \sigma_{i j}(h \nu) \mathrm{d} \nu   .
\end{equation}
We can get the function of the photoionization rate by integrating the photon frequency. The energy of the X-ray pulse can be obtained by integrating the intensity of space, time, and photon energy. We use a genetic algorithm to calibrate the peak flux and exponential coefficient of Kr atom at 17.50 keV photon energy, in which the peak flux is \num{1.61d12} photons/$\mu$m$^2$, the index coefficient is 1.85, so that the calculated value can better match the experimental value. For the parameters obtained by the genetic algorithm, a theoretical calculation of 1.75 keV, 1.90 keV, and 14.30 keV make our theoretical calculation results more reliable.

\subsection{Numerical method}  
The transition between electronic configurations can be well described by establishing the relevant rate equation between configurations. The equation form is as follows,

\begin{equation}
    \frac{\mathrm{d} n_{i}}{\mathrm{d} t}=\sum_{j \neq i}^{N} n_{j} R_{j i}-n_{i} \sum_{j \neq i}^{N} R_{i j} ,
\end{equation}
when solving the ionization dynamics of Kr atom at each photon energy, we use the Monte Carlo method, where $R_{i j}$ is the rate from state $i$ to state $j$, and $n_i$ is the ratio of the total number of events at energy level $i$ to the total number of events $N$. The total value of events depends on the parameters set by the Monte Carlo method. This method is often used to solve the ionization dynamics of atoms. It can give a better evolutionary relationship of atomic ion population when processing massive data. Moreover, when the electron number increases, the Monte Carlo method has been proved to be one of the most practical methods.\cite{son_monte_2012,rudek_ultra-efficient_2012} We ran 10$^6$ times to converge the results. We used the XATOM toolkit\cite{son_impact_2011} to calculate the photoionization cross-section, Auger decay, and fluorescence decay data of atoms. We calculated the photoionization cross-section, Auger rate, and fluorescence rate of Kr at 1.75 keV, 1.90 keV, 14.30 keV, and 17.50 keV\cite{krause_vacancy_1967,chernov_pseudopotential_2015} by using our own developed Monte Carlo program. A total of \num{8.19d4} configurations and \num{7.45d7} rate equations.

\section{Results and discussion}

\subsection{Ionization kinetics of Kr in 17.50 keV}

\begin{figure}
    \centering
    \includegraphics[width=\linewidth]{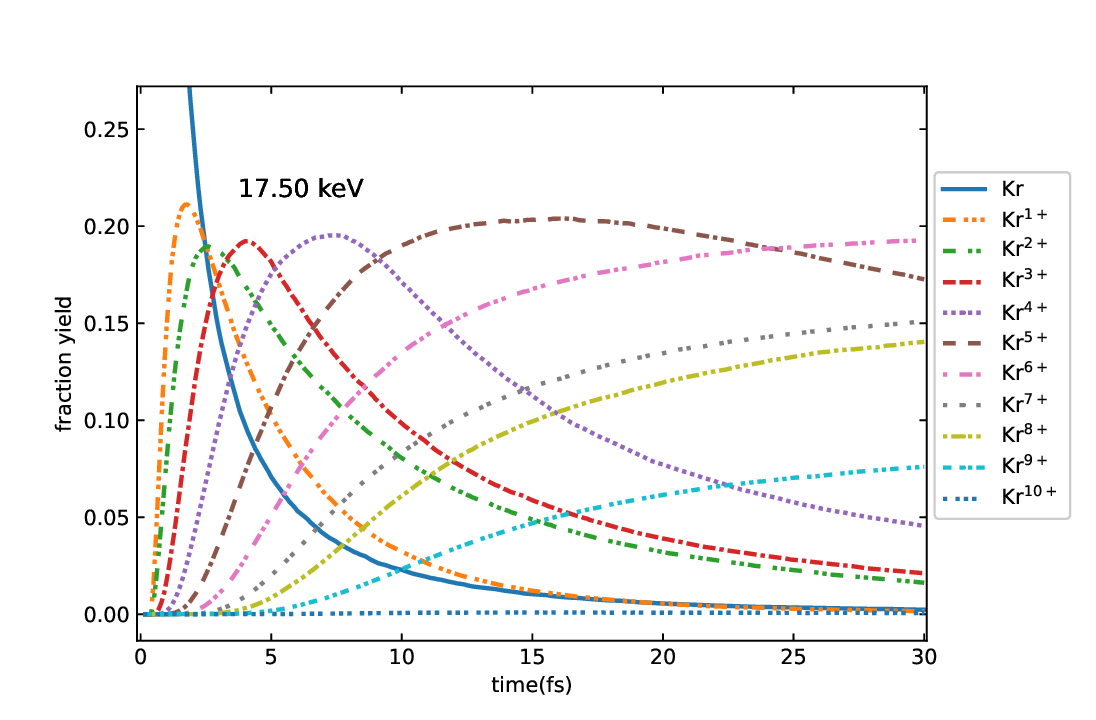}
    \caption{The population of Kr ions varies with time. Pulse width is 30 fs.}\label{fig:1}
\end{figure}

\begin{figure}
    \centering
    \includegraphics[width=\linewidth]{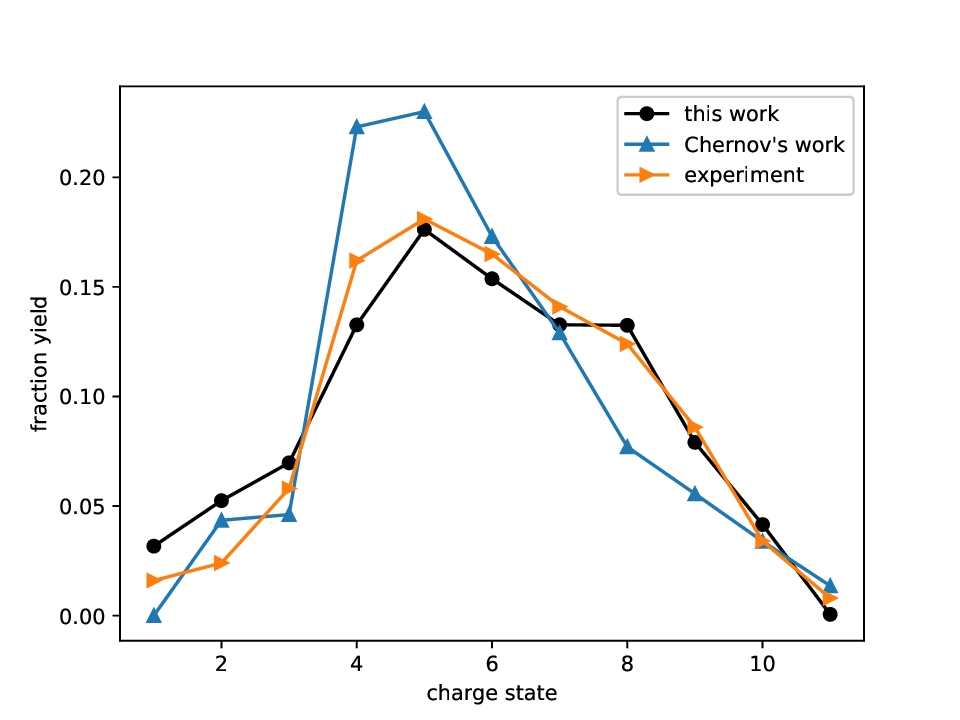}
    \caption{Charge state distribution of Kr at 17.50 keV photon energy compared with experimental data\cite{krause_vacancy_1967} and another theoretical result.\cite{chernov_pseudopotential_2015}}\label{fig:2}
\end{figure}

The time-dependent population obtained by the Monte Carlo method is about spatial distribution. Therefore, we integrate the population change of each ion over space to obtain its time-dependent population change in Fig. \ref{fig:1},

\begin{equation}
    n_i(t)=\int_{0}^{1} n_i(t, r) \mathrm{d} r  ,
\end{equation}
where $i = 1,2,3 \ldots 10$ is the population of the $i$th ion whose population changes with time. We made a statistical analysis of the calculation results of the Monte Carlo method and received the time-dependent change of the population of ions. The theoretical calculation results at different positions are integrated over time and space, and the population is obtained from each charge state compared with the experimental values, which are shown in Fig. \ref{fig:2}.

In comparison, our calculated value performed slightly worse in Kr$^{7+}$, and the other charge states with experimental values were better than the others.\cite{chernov_pseudopotential_2015} When compared with the experimental values, there were some differences between the Kr$^{1+}$, Kr$^{2+}$, and Kr$^{4+}$. The differences are possibly due to the relatively few physical processes considered, such as the addition of single photon multi-ionization, double Auger decay, and resonant excitation to the parameter fitting,\cite{li_effects_2022,ho_theoretical_2014} which can provide better agreement to the experimental results.

\subsection{Selection of CIPEs}

\begin{figure}
    \centering
    \includegraphics[width=\linewidth]{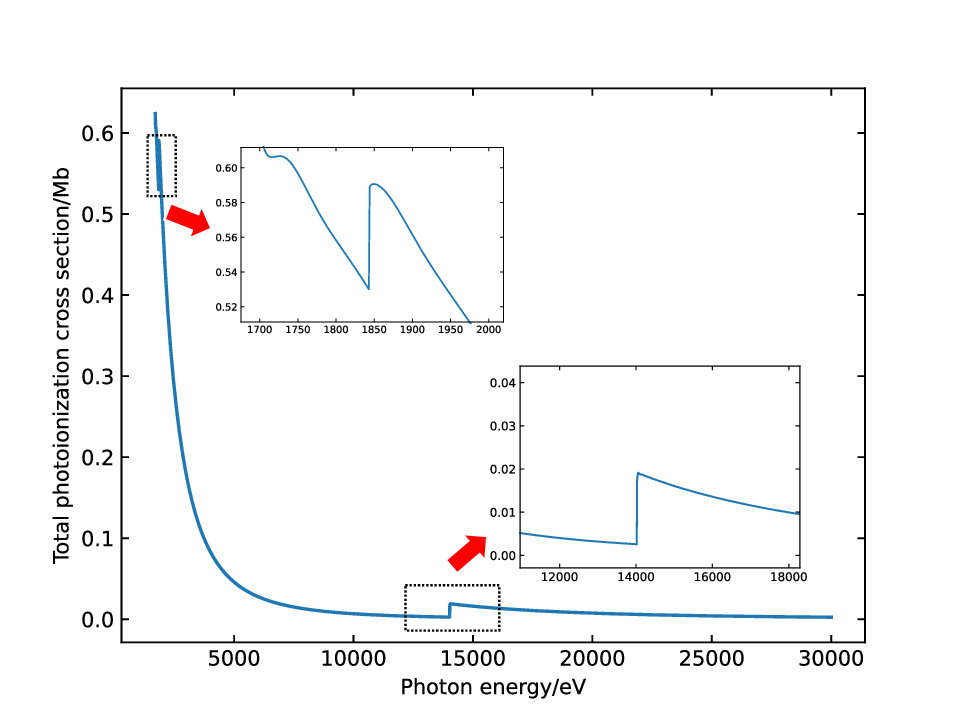}
    \caption{Total photoionization cross-section of Kr to photon energy (1.70--30.00 keV).}\label{fig:3}
\end{figure}

From Eq. \ref{eq:1}, we presented the relationship between total photoionization cross-section and photon energy. Fig. \ref{fig:3} indicates that the photoionization cross-section decreases with increasing photon energy. The total photoionization cross-section changes rapidly at photon energies near 1.70 keV, 1.87 keV, and 14.00 keV. When the photon energy approaches the ionization energy of each orbital, there will be an increase in the total ionization cross-section. When the orbital electron is ionized, the corresponding orbital energy increases. When the photon energy is smaller than the elevated orbital energy, the corresponding orbital will no longer undergo single-photon ionization, resulting in the enhancement of light transparency. We define the value of CIPE in the range of ground state orbital energy and ionized orbital energy. The interval of CIPEs for $2p$ is 1.67–1.76 keV, $2s$ is 1.84–1.92 keV, and $1s$ is 14.02–14.43 keV. We take the CIPEs of three different orbitals as 1.75 keV, 1.90 keV, and 14.30 keV, respectively. In XFEL with high photon energy, this assumption is reasonable. We call an atom a hollow-like atom (HLA) that ionizes an electron from the electron orbital with the ground state's most significant non-inner photoionization cross-section. Table \ref{tab:1} shows that hollow atoms and HLAs at three CIPEs and one experimental photon energy are compared.

\begin{table*}
\caption{\label{tab:1}%
The total photoionization cross-section includes the sum of 8
photoionization cross-section channels of $1s$--$4p$. The total rate includes
82 decay modes, 15 fluorescence decay rates and 67 Auger rates.
}
\begin{ruledtabular}
\begin{tabular}{dccc}
\multicolumn{1}{c}{\textrm{Photon energy (keV)}} & \textrm{Configuration} & \textrm{Total ionization cross-section (Mb)} &
\textrm{Total rates (au)}\\
\colrule
1.75    & \(1s^{2}2s^{2}2p^{6}\) & 0.597 & 0.000\\
        & \(1s^{2}2s^{2}2p^{5}\) & 0.114 & 0.045\\
1.90    & \(1s^{2}2s^{2}2p^{6}\) & 0.561 & 0.000\\
        & \(1s^{2}2s^{1}2p^{6}\) & 0.510 & 0.031\\
14.30   & \(1s^{2}2s^{2}2p^{6}\) & 0.018 & 0.000\\
        & \(1s^{1}2s^{2}2p^{6}\) & 0.003 & 0.102\\
17.50   & \(1s^{2}2s^{2}2p^{6}\) & 0.011 & 0.000\\
        & \(1s^{1}2s^{2}2p^{6}\) & 0.006 & 0.102\\
\end{tabular}
\end{ruledtabular}
\end{table*}

At 1.75 keV, the total photoionization cross-section decayed from 0.597 Mb to 0.114 Mb before and after ionization, and 14.30 keV decayed from 0.018 Mb to 0.003 Mb. However, the total ionization cross-section at 1.90 keV did not decay as expected because the ionization energy of $2s$ orbital was similar to that of $2p$ orbital. After ionization of $2s$ orbital electrons, the ionization energy of $2p$ orbital increased and approached 1.90 keV, resulting in the rise of the total ionization cross-section, which did not reach the expected attenuation value. 17.50 keV is a typical experimental parameter of photon energy from 1.7 keV to 30 keV. After the electron ionizes the inner orbital, the time and output of hollow atoms are less than the CIPE. It also shows that the light saturation effect of general photon energy is much lower than the CIPEs.

\subsection{Temporal and spatial distribution of Kr hollow atoms}

The generation of hollow atoms depends on time and photon flux,\cite{deng_charge-state_2019,feng_lei_2000_2017} however, in the actual process, we need to consider the distribution of spatial position as well. This paper assumes that the laser is a Gaussian beam; the incident photon beam satisfies the Gaussian distribution in spatial position. Through different position parameters, three-dimensional images of spatial and temporal populations of hollow atoms of 1.75 keV, 1.90 keV, 14.30 keV, and 17.50 keV are obtained, which shows the effect on hollow atoms when the position parameters change.

\begin{figure*} 
    \subfloat[1.75 keV]{
        \label{fig4a} 
        \includegraphics[width=0.49\linewidth]{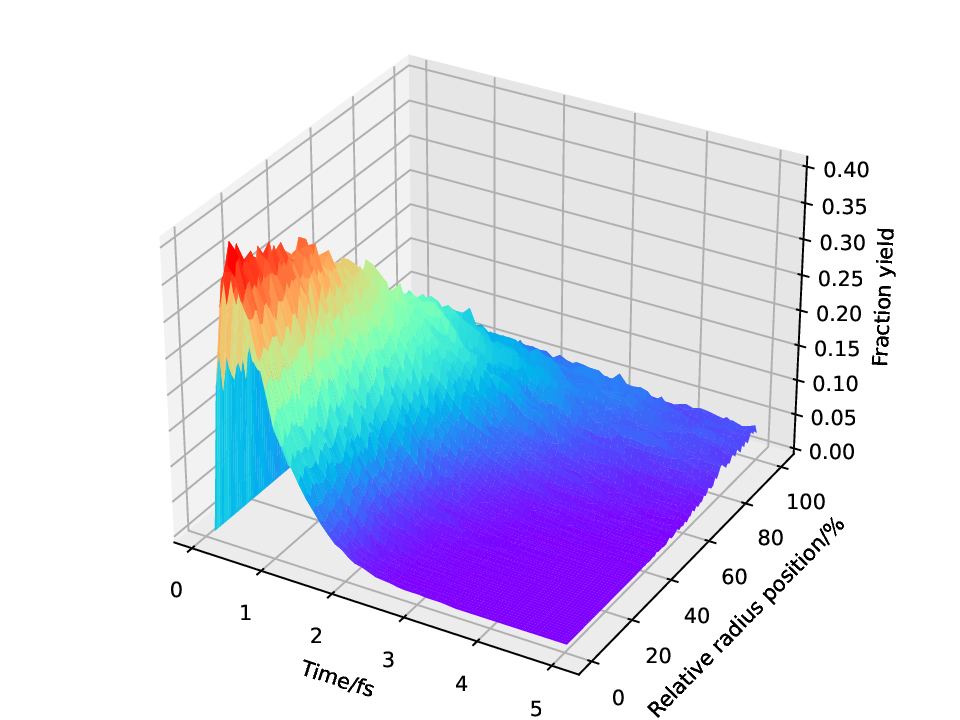}}
        \hfill
    \subfloat[1.90 keV]{
        \label{fig4b}
        \includegraphics[width=0.49\linewidth]{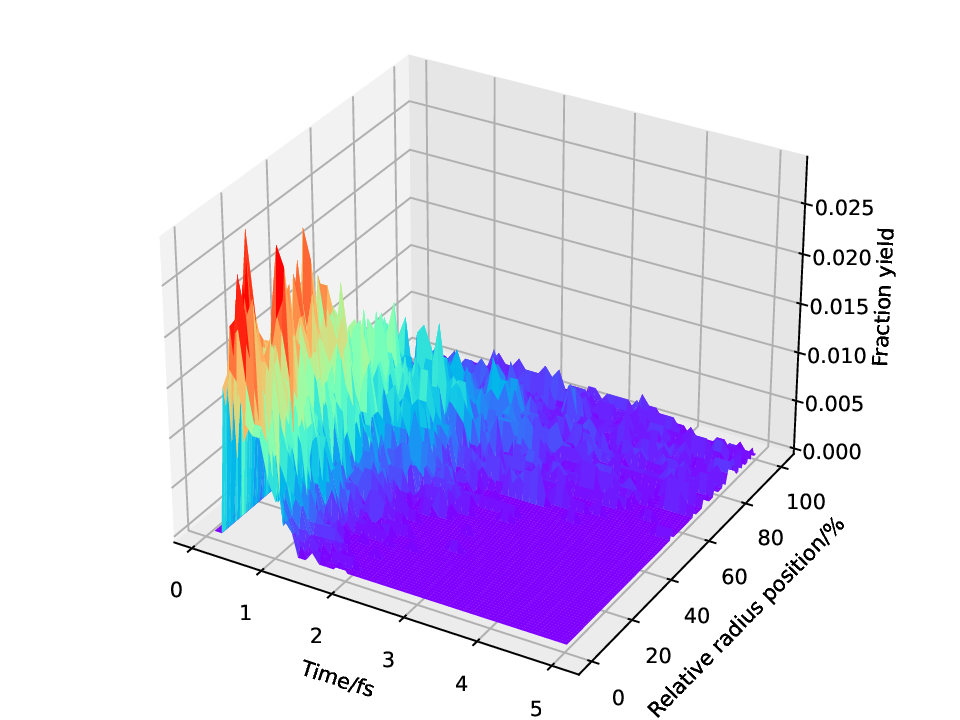}}
    \\
    \subfloat[14.30 keV]{
        \label{fig4c}
        \includegraphics[width=0.49\linewidth]{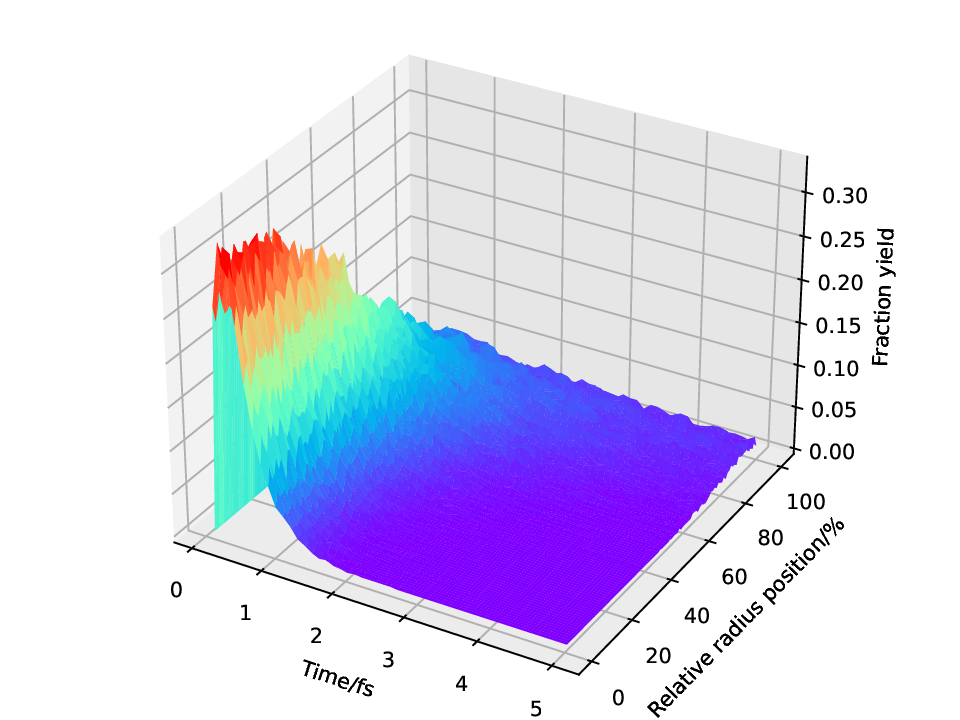}}
        \hfill
    \subfloat[17.50 keV]{
        \label{fig4d}
        \includegraphics[width=0.49\linewidth]{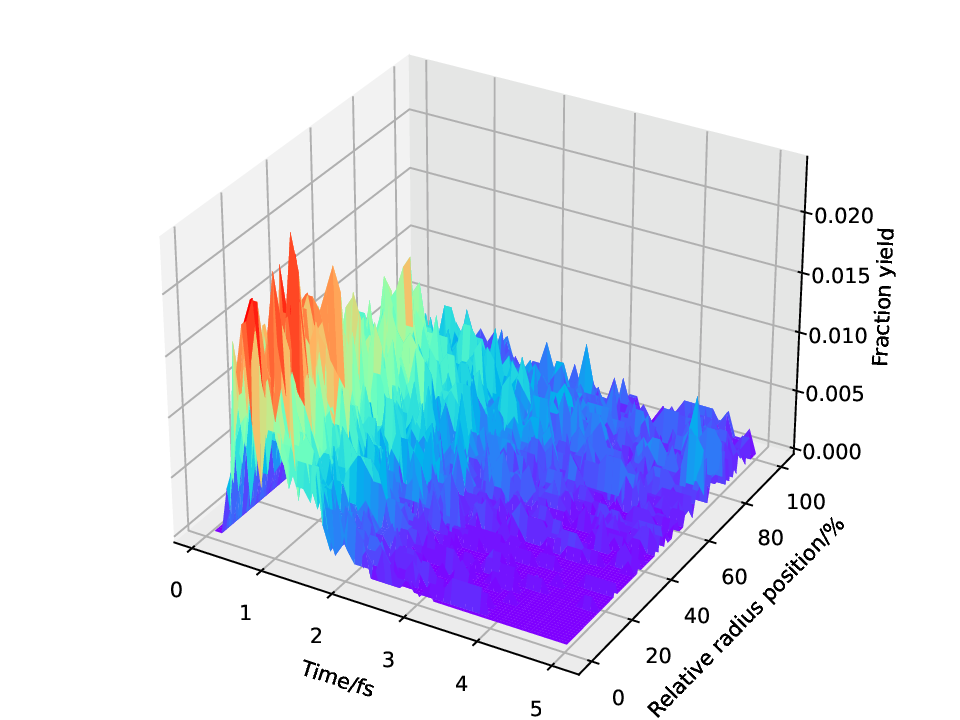}}
    \caption{Distribution diagram of hollow atoms and space. Time is from 0--5 fs, and position parameter is the relative radius position from 0--1 (100\%).}
    \label{fig:4} 
\end{figure*}

From Fig. \ref{fig:4}, we determined that 1.75 keV, 1.90 keV, 14.30 keV, and 17.50 keV reach the maximum average population of hollow atoms at 67\%, 35\%, 88\%, and 26\% percentage of the radius from the center point. Due to the characteristics of Gaussian distribution, with the distance from the center, the photon flux decreases, the number of hollow atoms decreases, and the change of population slow down. The maximum values of three-dimensional images are 40.5\%, 33.3\%, 2.9\%, and 2.4\% showing that the hollow atom effect produced by the CIPE of $1s$ and $2p$ is higher than the CIPE of 17.50 keV and $2s$, which is consistent with our prediction. By spatially integrating the image data, a more intuitive change of hollow atoms of Gaussian beams can be obtained.

\begin{figure}
    \centering
    \includegraphics[width=\linewidth]{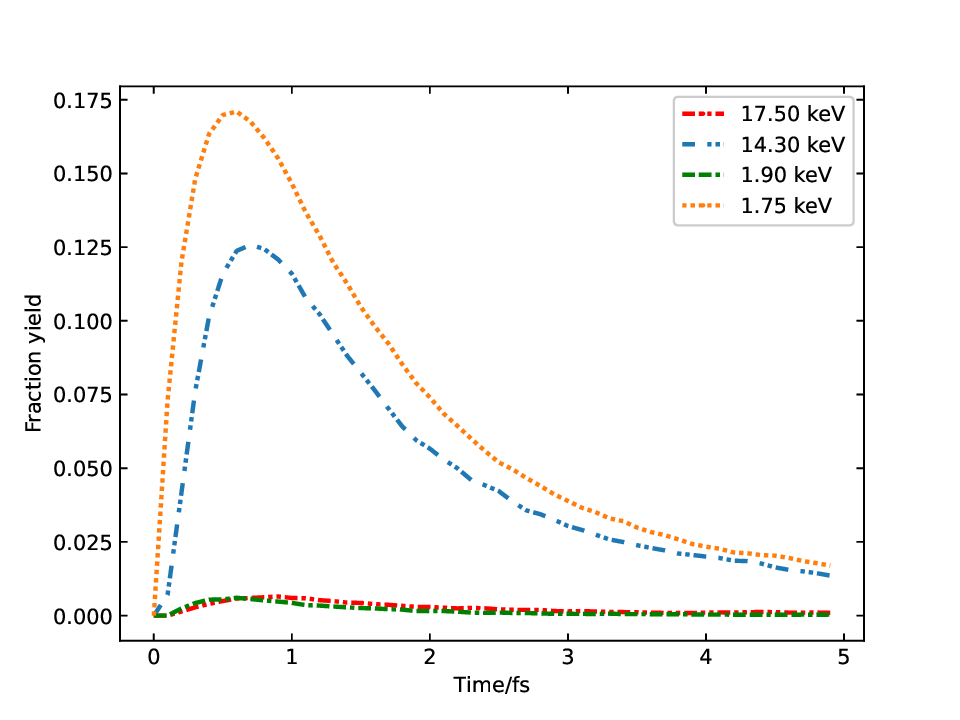}
    \caption{Time-dependent population change of hollow(-like) atoms. 1.75 keV and 1.90 keV are HLAs.}\label{fig:8}
\end{figure}

Fig. \ref{fig:8} shows that the populations of the two CIPEs of 1.75 keV and 14.30 keV are much higher than those of 1.90 keV and 17.50 keV photon energies, which reach the maximum at 0.6 fs and 0.7 fs. In the practical application of hollow atoms, the photon energy is the CIPE of the orbital with the largest angular momentum for a determined principal quantum number can be selected to make the diffraction imaging of material structure detection clearer.

\subsection{Ionization path diagram of Kr atom}

In over \num{2d4} configurations, studying the comparison of main ionization paths under each photon energy can help better understand the direction of hollow atoms. Therefore, we count and plot the main ionization paths at the four specific photon energies.

\begin{figure*}
    \centering
    \includegraphics[width=\linewidth]{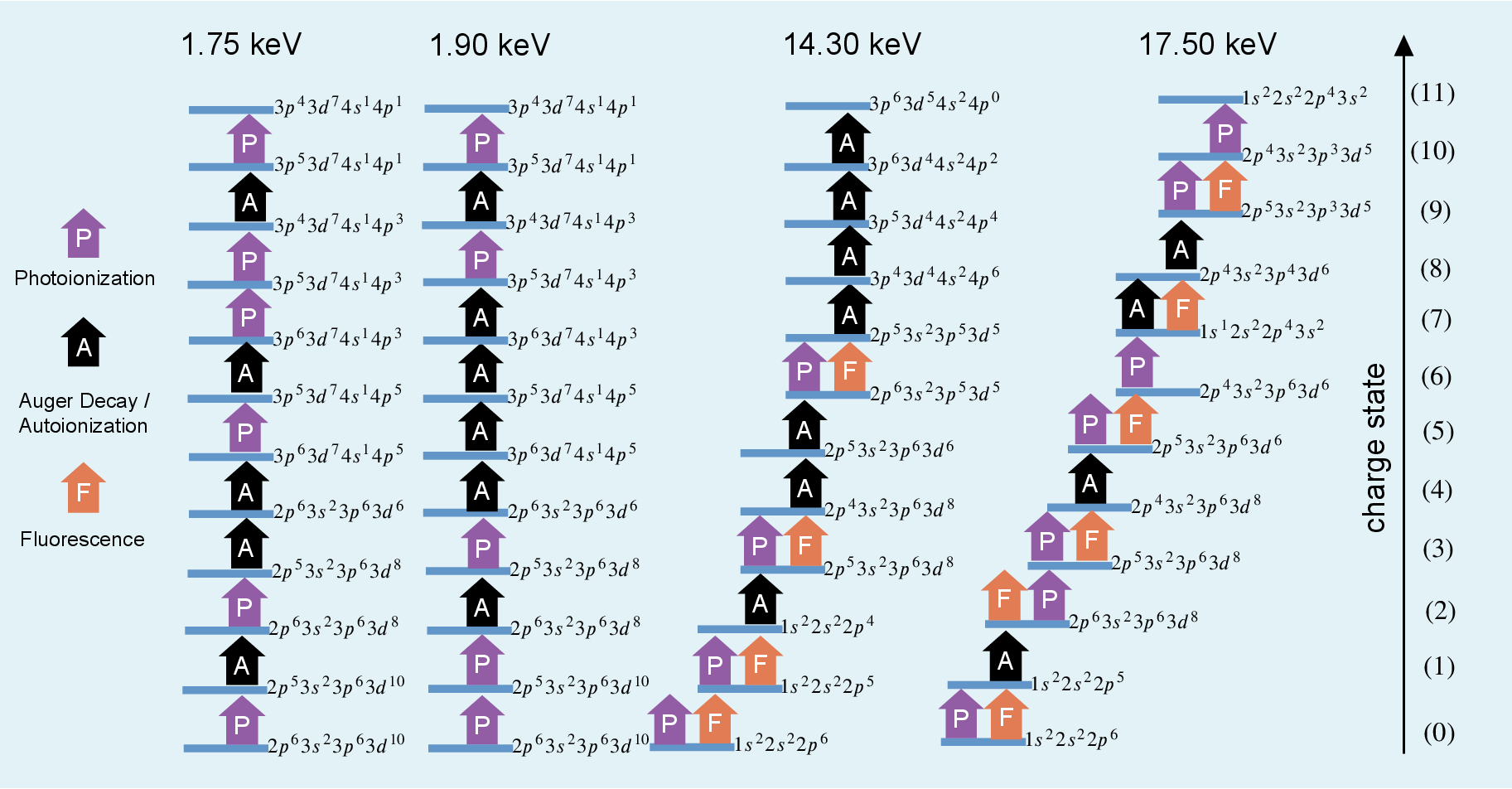}
    \caption{Diagram of the dominant photon energy ionization path. Configurations not shown and not mentioned are full-occupied electron shells; configurations not shown but mentioned inherit the configuration from the previous step. The charge state is shown in parentheses on the right side.}\label{fig:9}
\end{figure*}

Fig. \ref{fig:9} demonstrates that under the photon energy of 1.75 keV and 1.90 keV, single-photon photoionization and Auger decay account for all channels of ion generation, and fluorescence decay does not occur. At 14.30 keV, the ion generation mainly depends on the ion generation mode of inner photoionization and Auger decay. At 17.50 keV, the inner photoionization cross-section decreases, and the generation of ions mainly depends on the ion generation mode of inner photoionization plus fluorescence decay. The photon energy is close to the orbital energy at low photon energy, and the photoionization cross-section is large. Thus there will be no fluorescence decay in the primary ionization process.

On the other hand, the photon energy is no longer close to the orbital energy at high photon energy, resulting in fluorescence decay after inner photoionization. For HLAs, 1.75 keV is no longer ionized after ionizing $2p$ orbital electrons, and the main channel is Auger decay; therefore, the population of HLAs is high. For 1.90 keV, after ionizing the $2s$ orbital electrons, the photon energy is close to the $2p$ orbital energy. Then the $2p$ orbital ionization is mainly carried out, which is also the poor effect of producing hollow atoms in the CIPE. For hollow atoms, at 14.30 keV, because the photon energy is close to the inner orbital energy, and the ground state $1s$ photoionization cross-section and fluorescence decay ratio after ionization are higher than 17.50 keV at 14.30 keV, the hollow atom effect produced at 14.30 keV is more significant than 17.50 keV.

\section{Conclusion}

We used the Hartree-Fock-Slater model to calculate the total photoionization cross-section of the Kr atom in the range of 1.70--30.00 keV, calculated the ionization energy of each orbital of Kr, and finally determined three CIPEs of 1.75 keV, 1.90 keV, and 14.30 keV. We used the genetic algorithm to calibrate the parameters of the photon flux model in the results of calculating 17.50 keV photon energy Kr by the Monte Carlo method and calculated the dynamical evolution of hollow atoms of 1.75 keV, 1.90 keV, 14.30 keV, and 17.50 keV. The conclusion indicates that the number of hollow atoms of 1.75 keV and 14.30 keV is much higher than that of the other two hollow atoms with ionization energy, which provides an idea for selecting photon energy in the subsequent application of hollow atoms. The main ionization paths of Kr atoms at photon energies of 1.75 keV, 1.90 keV, 14.30 keV, and 17.50 keV were studied. We suggest that the main ionization paths rarely produce fluorescence decay at low photon energy. In contrast, the generation of ions at high photon energy depends on the ionization mode of inner ionization and fluorescence decay, which provides an idea for understanding the ionization mode of high-$Z$ atoms at different photon energies.

\begin{acknowledgments}
Project was supported by the Fundamental Research Funds for the Central Universities (Grant No. 10822041A2038). The authors would like to thank all the people who have been helping us in the theoretical calculation; particularly, our thanks go to Prof. Robin Santra, Dr. Yongjun Li, and Prof. Jianmin Yuan.
\end{acknowledgments}

\bibliographystyle{apsrev4-2}
\bibliography{yjxbib}

\end{document}